\documentclass[aps,prd,showpacs,amssymb,amsmath,preprint]{revtex4}
\usepackage{graphicx}
\usepackage{bm}

\begin{document}
\title{Decoherence induced by fluctuating boundaries}
\author{V. A. \surname{De Lorenci}}
 \email{delorenci@unifei.edu.br}
\author{L. H. \surname{Ford}}
 \email{ford@cosmos.phy.tufts.edu}
\affiliation{$\mbox{}^{*}$Instituto de Ci\^encias Exatas,
Universidade Federal de Itajub\'a, Itajub\'a, MG 37500-903, Brazil} 
\affiliation{$\mbox{}^\dag$Institute of Cosmology, Department of Physics and Astronomy,
Tufts University, Medford, MA 02155, USA}

\date{\today}

\begin{abstract}
The effects of fluctuating boundaries on a superposition state
of a quantum particle in a box is studied. We consider a model 
in one space dimension in which the initial state is a coherent
superposition of two energy eigenstates. The locations of the
walls of the box are assumed to undergo small fluctuation with
a Gaussian probability distribution. The spatial probability density 
of the particle contains an interference term, which is found to decay
in time due to the boundary fluctuations. At late times, this term
vanishes and the quantum coherence is lost. The system is now
described by a density matrix rather than a pure quantum state.
This model gives a simple illustration of  how environment-induced 
decoherence can take place in quantum systems.  It can also serve as an 
analog model for the effects of spacetime geometry fluctuations on
quantum systems. 
\end{abstract}

%

\maketitle

\baselineskip=14pt
\section{Introduction}
The theoretical study of quantum systems is usually performed by assuming
that for all practical purposes they can be considered as isolated. However,
realistic physical systems are never isolated. Instead, they are always immersed
in an environment that continuously interacts with them~\cite{zeh1970}. 
As a result of such interactions, quantum superpositions tend to be suppressed and 
are not usually found at the macroscopic level. Exceptions occur in the phenomenon
of superfluidity, and possibly in gravitational wave detection technology, where  
quantum effects can arise on macroscopic scales~\cite{braginsky1980,caves1980}.
The mechanism behind the loss of coherence between the components 
states in a quantum superposition has been called environment-induced
decoherence (decoherence for short)~\cite{zurek1981,zurek1982,zurek1991}, and 
it has been accepted as the mechanism responsible to the emergence of the classical world from the 
quantum physics~\cite{joos1985,zurek1986}. Decoherence has been subjected to intense 
investigation 
in recent decades and plays important roles in several areas of physics, including 
quantum field theory, condensed matter, quantum optics, quantum-chromo-dynamics
and cosmology (for  reviews, see Refs.~\cite{paz2001,zurek2003,joss2003,schlosshauer2007} 
and references therein).

Imposition of fluctuating boundary conditions on quantum fields in certain
physical configurations was  proposed in Ref.~\cite{ford1998}.  The motivation was
to address the problem of infinite energy density appearing when fixed
boundaries are assumed~\cite{fulling89}. It was found that boundary fluctuations
can lead to finite energy densities and a natural description of surface energy densities. 

In this paper, we study a simple model of a particle described by a superposition
of two quantum states inside a potential well in one space dimension. 
Fluctuating boundaries are 
introduced as a mechanism modelling environment interactions. 
The averaged probability density is derived and its time evolution is investigated.  
It is shown that the net effect  of the fluctuations is to suppress the interference
terms between the component states of the particle wave function. 
As time goes on, the resulting probability density reduces to the case corresponding to
a statistical mixture of states. The results indicate that this model can be understood
as an example of decoherence. The effects of the fluctuating boundaries are 
similar to the effects of fluctuating or time dependent electromagnetic fields
on quantum charged particles~\cite{HF04}.

It is well known that boundary conditions in flat spacetime can
model certain aspects of the effects of spacetime geometry or gravity on quantum 
fields. For example, the vacuum energy of quantum fields in an Einstein universe
can be derived as a Casimir energy, using the same techniques used to find
Casimir energies in cavities in flat spacetime~\cite{F75}. 
This suggests that fluctuating boundaries 
might be used as analog models for the effects of fluctuations of the gravitational
field or of the background spacetime geometry. We will not pursue this idea in detail
in the present paper, but leave it as a topic for future study.

The simple quantum mechanical system of a particle in a superposition 
of states of a potential well is developed in the following section. In Sec. III, 
we model the interaction between the system and its environment by means of 
fluctuating boundaries, described by a normal probability distribution. 
The averaged probability density of the particle is derived and its time evolution
is discussed. A numerical example is presented at the end of this section.
Final remarks are outlined in Sec. IV, including a short discussion on 
the extension of the model to quantum systems in three space dimensions and
to the case of electromagnetic waves in a cavity.

\section{The Model}
We begin with a non-relativistic particle of mass $m$ confined in an 
infinite potential well whose width is denoted by $a$. For the sake of simplicity we
restrict our analysis to the one-dimensional case. 
Suppose that the normalized state of the particle in an arbitrary
time $t$ is given by a superposition of the first two available states as
\begin{equation}
\psi(x,t) = \frac{1}{\sqrt{2}}\left[\psi_1(x,t) + \psi_2(x,t)\right].
\label{5}
\end{equation}
The eigenfunctions $\psi_n$ ($n=1,2$) are independent solutions of the 
Schr\"odinger equation under the conditions that these functions 
vanish on the boundaries at $x=0$ and $x=a$, yielding 
\begin{equation}
\psi_n(x,t) = \sqrt{\frac{2}{a}}\,\sin\left(\frac{n \pi x}{a}\right) \;{\rm e}^{-i\omega_n t},
\label{1}
\end{equation}
with $\omega_n = n^2 \pi^2 \hbar / 2 m a^2$. Each one of these eigenfunctions
describes a stationary state, as the corresponding probability density $|\psi_n|^2$ 
is time-independent. However, time evolution occurs when the
particle is governed by a linear superposition of $\psi_n(x,t)$, such as that 
given in Eq. (\ref{5}). The probability density can be obtained as
\begin{eqnarray}
|\psi(x,t)|^2 = \frac{1}{2}\left(|\psi_1|^2 + |\psi_2|^2\right) 
+ |\psi_1||\psi_2|\cos\omega t,
\label{11}
\end{eqnarray}
where we define $\omega = \omega_2 - \omega_1$, which represents the corresponding 
Bohr angular frequency of the system.
As we see, the time evolution of $|\psi(x,t)|^2$ is exclusively governed 
by the interference term between $\psi_1$ and $\psi_2$. In fact, $|\psi(x,t)|^2$
oscillates between its maximum $(|\psi_1|+|\psi_2|)^2/2$ and minimum
$(|\psi_1|-|\psi_2|)^2/2$ values, as depicted in the down inset frame in 
Fig. \ref{fig1}, for a particular numerical model. 

\section{Probability density with fluctuating boundaries}
Now we wish to investigate the behavior of this quantum system when interaction
with the environment takes place. In order to model the interaction
between the system and its environment, we allow the positions of the
physical boundaries to fluctuate under the influence of an external
noise. This can simply be implemented by allowing the width 
parameter $a$ to undergo fluctuations around a mean value $\bar a$ \cite{ford1998}. 
We set $a = \bar{a}(1+\varepsilon)$, where the dimensionless parameter $\varepsilon$
is described by a Gaussian distribution as
\begin{equation}
f(\varepsilon) = \sqrt{\frac{\theta}{\pi}}\;{\rm e}^{-\theta \varepsilon^2},
\label{13}
\end{equation}
with $\theta$ related to the width $\sigma$ of the distribution by means
of $\sigma^2 = 1/2\theta$. The mean value of a arbitrary function $G(\varepsilon)$
over $\varepsilon$ is a linear operation defined by
\begin{equation}
\left<G\right>  = \int_{-\infty}^{\infty}G(\varepsilon)f(\varepsilon)d\varepsilon.
\label{15}
\end{equation}
Notice that  $\left<\varepsilon^2\right> -\left<\varepsilon\right>^2 = \sigma^2$, 
the mean squared fluctuation of $\varepsilon$.

There does not seem to be a meaning to averaging the wave function $\psi(x,t)$, as it is 
not directly observable. 
(See, however, Ref.~\cite{lundeen2011} for a discussion of the possibility of measuring a wave 
function in the context of a weak measurements approach.) 
Here we are interested in studying averaged values of observable quantities. Particularly,
the modulus squared of the particle wave function, Eq.~(\ref{11}), represents the probability 
density associated with the  position the particle .  
The average over width fluctuations of this quantity can be calculated by using
Eqs. (\ref{11}) and (\ref{15}), yielding
\begin{eqnarray}
\left<|\psi|^2\right> = \frac{1}{2}\left(\left<|\psi_1|^2\right> + \left<|\psi_2|^2\right>\right) 
+ \left<|\psi_1||\psi_2|\cos\omega t\right>.
\label{22}
\end{eqnarray}
Here the angular bracket refers to the average over positions of the boundaries, which
was defined in Eq.~(\ref{15}).
In what follows, each term appearing in the above equation will be considered separately.

The two first terms appearing in Eq.~(\ref{22}) can be  expressed, using Eqs.~(\ref{1}) and 
(\ref{15}), as  
\begin{equation}
\left<|\psi_n|^2\right> =  \sqrt{\frac{4\theta}{\pi}} \int_{-\infty}^{\infty} 
\frac{{\rm e}^{-\theta \varepsilon^2}}{\bar{a}(1+\varepsilon)} \sin^2 \left[\frac{n\pi x}
{\bar{a}(1+\varepsilon)} \right] \, d\varepsilon \,,
\label{23}
\end{equation}
where $n = 1,2$.
We assume a narrow distribution, $\sigma \ll 1$, which means that only small values
of $\varepsilon$ contribute in Eq.~(\ref{23}). First use the identity 
$\sin^2 u = [1- {\rm Re( e}^{2iu})]/2$. Next Taylor expand to first order in $\varepsilon$
inside the exponential, but only to zeroth order otherwise. The integrations may be
performed using the identity
\begin{equation}
\int_{-\infty}^{\infty} e^{\pm iZ\varepsilon -\theta\varepsilon^2} d\varepsilon
= \sqrt{\frac{\pi}{\theta}} e^{-Z^2/4\theta} \,.
\label{29}
\end{equation}
The result is 
\begin{equation}
\left<|\psi_n|^2\right>  \approx  \frac{1}{\bar{a}} - \frac{1}{\bar{a}} 
\cos\left(\frac{2n\pi x}{\bar a}\right) \;
\exp \left(\frac{-2 n^2\pi^2\sigma^2x^2}{\bar{a}^2}\right) \,.
\label{31}
\end{equation}
Further, as the particle is confined ($0\le x \le \bar a$), and the small $\sigma$
 approximation, $\pi\sigma x/\bar a \ll 1$, is assumed, we have
\begin{equation}
\left<|\psi_n|^2\right>\, \approx 
\frac{2}{\bar a}\sin^2\left(\frac{n \pi x}{\bar a}\right)  =  |\psi_n|^2.
\label{34}
\end{equation}
Thus, the probability density associated with the energy eigenstates
 does not   change significantly when
boundary fluctuations are introduced. In other words, if the particle is initially
in an energy eigenstate state $\psi_n(x,t)$, it will remain in this state and its
probability density will not undergo appreciable time evolution
when small boundary fluctuations are present.

Next, we consider the last term in Eq.~(\ref{22}), which describes the interference
effects occurring in the system. From Eqs.~(\ref{1}) 
and (\ref{15}), we obtain that 
\begin{eqnarray}
\left<|\psi_1||\psi_2|\cos\omega t\right> &=& 
\frac{1}{4\bar a}\sqrt{\frac{\theta}{\pi}}\left(A_{1}^{+}
+A_{-1}^{+}-A_{3}^{+}-A_{-3}^{+} 
\right. \nonumber\\ 
&& \left.+A_{1}^{-}+A_{-1}^{-}-A_{3}^{-}-A_{-3}^{-}\right), 
\label{35} 
\end{eqnarray}
where $A_q^{\pm}$ is defined by
\begin{eqnarray}
A_q^{\pm}=  \int_{-\infty}^{\infty} \frac{1}{1+\varepsilon}
\exp\left[\frac{q i \pi x}{\bar a (1+\varepsilon)} \pm 
\frac{i\bar{\omega} t}{(1+\varepsilon)^2} - \theta\varepsilon^2\right]\, d\varepsilon ,
\label{36}
\end{eqnarray}
and $\bar\omega  = \omega_{2}(\bar a) - \omega_{1}(\bar a) = 3\hbar\pi^2/2m\bar{a}^2$.
Proceeding as before, assuming $\sigma \ll 1$ and using Eq.~(\ref{29}), we obtain 
\begin{eqnarray}
A_q^{\pm} \approx \sqrt{\frac{\pi}{\theta}}
\exp\left[\frac{iq\pi}{\bar a}
\left(x\pm\frac{\bar a \bar\omega t}{q\pi}\right)\right]
\nonumber\\
\times\exp\left[\frac{-q^2\pi^2}{4\theta\bar{a}^2}
\left(x\pm\frac{2\bar a \bar\omega t}{q\pi}\right)^2\right] \,.
\label{39}
\end{eqnarray}

After a finite time, $t \agt 1/\bar\omega$, the $x$ dependence in the real exponential 
becomes unimportant, and $A_q^{\pm}$ can be approximated by 
\begin{eqnarray}
A_q^{\pm}\approx \sqrt{\frac{\pi}{\theta}}
\exp\left[\frac{iq\pi}{\bar a}
\left(x\pm\frac{\bar a \bar\omega t}{q\pi}\right)\right]
e^{-\Gamma t^2},
\label{43}
\end{eqnarray}
where $\Gamma = 2\bar\omega^2\sigma^2$. 
Using this result in Eq.~(\ref{35}), we obtain
\begin{eqnarray}
\left<|\psi_1||\psi_2|\cos\omega t\right> &=& 
\frac{1}{\bar a}\left[\cos\left(\frac{\pi x}{\bar a}\right)
-\cos\left(\frac{3\pi x}{\bar a}\right)\right] 
\nonumber \\
&&\times\cos(\bar\omega t) \;  e^{-\Gamma t^2}.
\label{44} 
\end{eqnarray}
As one can see this term describes oscillations modulated by a factor which decays
exponentially in squared time.
 The time scale for the onset of this  decay is 
 \begin{equation}
 t_o = \frac{1}{\bar\omega}\;.
 \end{equation}
 In the
case of an electron in a potential well with $\bar a \approx 1 \mbox{\AA}$, this time is 
of order  $t_o \approx10^{-17}\, s$. However, once the decay begins, the 
characteristic decay time is
\begin{equation}
 t_d = \frac{1}{\sqrt{\Gamma}} = \frac{t_o}{\sqrt{2}\, \sigma}\,,
 \end{equation}
which is longer by a factor of about $1/\sigma$.

Combining  the results in Eqs.~(\ref{34}) and (\ref{44}) with Eq.~(\ref{22}), we
find that the average $\left<|\psi(x,t)|^2\right>$ becomes
\begin{eqnarray}
\!\!\!\!\!\!\left<|\psi|^2\right>&\approx& \!\frac{1}{2}\left(|\psi_1|^2 + |\psi_2|^2\right)
\nonumber \\
&&+\frac{1}{\bar a}\!\left[\cos\left(\frac{\pi x}{\bar a}\right)-
\cos\left(\frac{3\pi x}{\bar a}\right)  \!\right] 
\!\cos(\bar\omega t) e^{-\Gamma t^2}.
\label{45}
\end{eqnarray}
As time passes, the last term in the above equation  falls to zero for $t \gg t_d$. 
Thus, the net effect of the fluctuations is to kill the interference
term. Neglecting the last term in Eq. (\ref{45}) we obtain that 
\begin{eqnarray}
\left<|\psi|^2\right> = \frac{1}{2}\left(|\psi_1|^2 + |\psi_2|^2\right), 
\label{46}
\end{eqnarray}
which corresponds to a weighted sum of probabilities as occurs when a 
statistical mixture of states is considered. 

The magnitude of the parameter $\sigma$ depends upon the source of the
boundary fluctuations. However, it is useful to make a simple model in which
the boundaries are massive particles in a harmonic potential well. The 
ground state of a quantum harmonic oscillator is described by a Gaussian
with characteristic width
\begin{equation}
\Delta x = \sqrt{\frac{\hbar}{2M\, \omega_0}} \,,
\end{equation}
where $M$ is the mass of the boundary, and $\omega_0$ is the angular frequency
of the harmonic oscillator. Plausible choices for these parameters might be a nuclear
mass and the vibrational frequency of a nucleus in a molecule or a crystal
lattice:
\begin{equation}
\Delta x \approx 0.01\, \text{\AA} \, \left( \frac{30\, {\rm amu}}{M} \right)^\frac{1}{2}
\;   \left(\frac{10^{15}\, {\rm s}^{-1}}{\omega_0} \right)^\frac{1}{2} \,.
\end{equation}
With these choices, and assuming $\bar a \approx 1 \mbox{\AA}$, 
we have $\sigma \approx 10^{-2}$.

\begin{figure}[!hbt]
\leavevmode
\centering
\includegraphics[scale = 1.0]{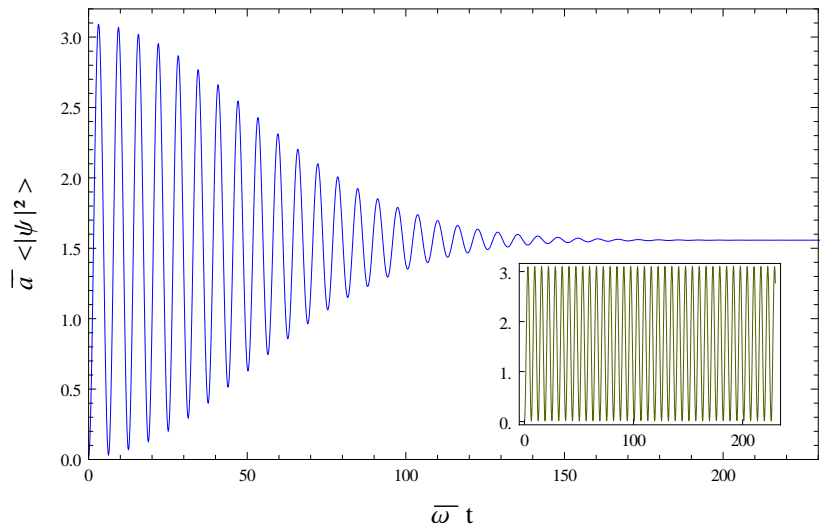}
\caption{{\small\sf (color online). 
The figure shows the behavior of the probability density associated with 
the  state defined by Eq. (\ref{5}) when fluctuating boundaries 
are assumed. As time goes on, interference effects between 
the component states of $\psi(x,t)$ are suppressed.  
For fixed boundaries, no suppression of interference 
is found, as shown in the down inset frame. 
We set $x/\bar{a}=0.7$ and $\sigma=0.01$.}}
\label{fig1}
\end{figure}

The behavior described by Eq.~(\ref{45}) can be illustrated by a numerical example. 
For instance, let us study the time evolution of the averaged probability 
density when fluctuating boundaries are considered in a particular model
with $x/\bar a = 0.7$ and $\sigma = 0.01$. In this case Eq.~(\ref{22}) 
can be integrated numerically. The result is depicted in Fig.~\ref{fig1}. 
Alternatively we could have used the approximate solution given by Eq.~(\ref{45}),
which leads to an identical graph, confirming the approximation used in
obtaining Eq.~(\ref{45}).   
As the figure shows clearly, the probability density $\left<|\psi|^2\right>$ oscillates
around the mean value $(\left<|\psi_1|^2\right> + \left<|\psi_2|^2\right>)/2$,
converging to this value when $t \gg t_d$. As anticipated, the net effect of 
 the fluctuating boundaries of the potential well is to 
kill the interference effect between the two state components $\psi_1$ and $\psi_2$.  
Finally, when no fluctuations are present, the usual stationary solution holds, as
illustrated by the down inset frame in Fig. \ref{fig1}.

In order to have an estimate, consider again the case of an electron in a potential
well with $\bar a\sim 1\mbox{\AA}$. As shown in Fig.~\ref{fig1} the oscillations 
in $\left<|\psi|^2\right>$
are completely suppressed when $\bar\omega t = 200$, that is, $10^{-14}$ seconds after
the boundary fluctuations are turned on. 

\section{Discussion}
We have studied a simple one dimensional model of a quantum particle confined
by fluctuating boundaries. A coherent superposition of energy eigenstates is quickly
converted into a statistical mixture, with the interference term being damped in time 
as $e^{-\Gamma t^2}$. This seems to be a simple example of decoherence in quantum 
systems. This behavior can be understood as a loss of phase coherence. A quantum
particle in a box can be viewed as undergoing repeated reflections from the box walls.
The uncertainty in position of the walls leads to uncertainty in the phase of the reflected
wave, which accumulates in time. 

Although we considered only one space dimension for simplicity, the same features
are to be expected for a quantum particle in a three dimensional box with fluctuating
walls. The crucial feature is the appearance of an oscillating interference term, such as 
that in Eq.~(\ref{22}), whose frequency depends upon the position of the boundaries,
which will also be the case in three dimensions. It is not necessary that all of the boundaries
undergo fluctuations. Even if only one boundary is subject to fluctuations, while
all others are fixed, this will introduce a dependence on the fluctuation parameter
in the time dependence of the interference term. When averages over the fluctuations
are taken, a decaying exponential in squared time will result, just as in Eq.~(\ref{44}).

The same comments apply to standing electromagnetic waves in a cavity with
fluctuating boundaries. We could average the energy density of such waves
over the position of the boundary, and obtain essentially the same results as
for a quantum particle. If a single normal mode of the cavity is excited, then the
energy density will not be significantly altered by small fluctuations, as is the case for energy
eigenstates of a quantum particle. However, a coherent superposition of normal
modes will have an interference term in its energy density which will decay in
time in the presence of boundary fluctuations. Note that this effect has nothing
to do with the damping in the electric field oscillations due to power losses in 
a cavity. Rather it  reflects a loss of the definite phase relation between the superposed
modes.

In summary, the effects of boundaries with fluctuating positions is a simple model for
decoherence in a quantum system. The effects of fluctuating boundaries can model
external influences such as fluctuating electromagnetic or gravitational fields.

\begin{acknowledgments}
This work was partially supported by the Brazilian research agencies CNPq and
FAPEMIG and by the National Science Foundation under Grant  PHY-0855360.
\end{acknowledgments}

\end{document}